\documentclass{svproc}
\usepackage{url}
\usepackage{graphicx}
\usepackage{amsfonts,amssymb}   % ADDED PACKAGE FOR MATH SYMBOLS
\usepackage{lineno}             % ADDED PACKAGE FOR LINE NUMBERS
\usepackage{xcolor}             % ADDED PACKAGE FOR COLOURS

\begin{document}
\mainmatter              % start of a contribution
\title{Community detection by resistance distance: automation and benchmark testing}
\titlerunning{Community detection by resistance distance}  % abbreviated title (for running head)
\author{Juan Gancio\inst{1} and Nicol{\'a}s Rubido\inst{2,1}}
\authorrunning{J. Gancio \& N. Rubido} % abbreviated author list (for running head)
\tocauthor{Juan Gancio and Nicol{\'a}s Rubido}
\institute{1 -Universidad de la Rep{\'u}blica, Instituto de F{\'i}sica de Facultad de Ciencias, Igu{\'a} 4225, Montevideo 11400, Uruguay \\
2 - University of Aberdeen, Aberdeen Biomedical Imaging Centre, AB25 2ZG Aberdeen, United Kingdom}
%,\\
%\email{I.Ekeland@princeton.edu},\\ WWW home page:
%\texttt{http://users/\homedir iekeland/web/welcome.html}
%\and
%Universit\'{e} de Paris-Sud, Laboratoire d'Analyse Num\'{e}rique, B\^{a}timent 425,\\F-91405 Orsay Cedex, France}

\maketitle              % typeset the title of the contribution
%%%%%%%%%%%%%%%%%%%%%%%%%%%%%%%%%%%%%%%%%%%%%%%%%%%%%%%%%%%%%%%%%%%
\begin{abstract}
Heterogeneity characterises real-world networks, where nodes show a broad range of different topological features. However, nodes also tend to organise into communities -- subsets of nodes that are sparsely inter-connected but are densely intra-connected (more than the network's average connectivity). This means that nodes belonging to the same community are close to each other by some distance measure, such as the resistance distance, which is the effective distance between any pair of nodes considering all possible paths. In this work, we present automation (i.e., unsupervised) and missing accuracy tests for a recently proposed semi-supervised community detection algorithm based on the resistance distance. The accuracy testing involves quantifying our algorithm's performance in terms of recovering known synthetic communities from benchmark networks, where we present results for Girvan-Newman and  Lancichinetti-Fortunato-Radicchi networks. Our findings show that our algorithm falls into the class of accurate performers.

\end{abstract}
%\linenumbers
%%%%%%%%%%%%%%%%%%%%%%%%%%%%%%%%%%%%%%%%%%%%%%%%%%%%%%%%%%%%%%%%%%%
%%%%%%%%%%%%%%%%%%%%%%%%%%%%%%%%%%%%%%%%%%%%%%%%%%%%%%%%%%%%%%%%%%%
\section{Introduction}
A network is an abstract model of the inter-relationships (links) between the elements (nodes) of a data-set \cite{BOCCALETTI2006175}, which can result in extremely complex structure. However, real-world networks also show the presence of communities; that is, densely connected groups of nodes with sparse inter-group connections \cite{FORTUNATO201075}. The ability to detect these community structures has many practical applications \cite{wasserman1994social}, mainly, because it simplifies the network analysis to clustered subgroups.

Because we lack a strict definition of community, the detection problem may present different solutions (particularly when dealing with a definition that uses macroscopic quantities such as the density of links), leading to the development of many methods with varying degree of success \cite{gansner1998improved,radicchi2004defining,reichardt2004detecting,guimera2004modularity,medus2005detection,palla2005uncovering,Newman8577,Newman.PRE.74.036104,alves2007,noack2009modularity,PhysRevE.81.046106,quiles2016dynamical,calatayud2019exploring,zhang_and_bu_2019}. A community detection method success requires testing its accuracy on networks where the community structure is known \cite{girvan2002community,danon2005comparing,fortunato_benchmark,gouvea2021force}. This can be achieved by selecting benchmark networks, such as the synthetically generated by Girvan and Newman (GN) \cite{girvan2002community} or Lancichinetti, Fortunato, and Radicchi (LFR) \cite{fortunato_benchmark} benchmark models.
%brought into consideration the need for a new benchmark of networks for the evaluation of the different algorithms that were developed for community detection. Previously the most common used benchmark was the class of networks proposed by Girvan and Newman \cite{girvan2002community}. This class of networks is usually set with 128 nodes, divided in 4 groups of 32 nodes each. Lancichinetti et al, criticize this class of networks in three points: every node have almost the same degree, the communities are of the same size, and the network is small. They argue that the degree distribution of real world networks follows a power law, that this is also the case for the size distribution of the communities, and that the network size is insufficient to compare the performance of modern algorithms, that can handle networks of $10^6$ nodes. 

Recently, Zhang and Bu (ZB) \cite{zhang_and_bu_2019} proposed a new method for community detection based on the resistance distance. The resistance distance includes more information than the shortest paths, since it also considers every different possible path between any two nodes, weighing them as parallel paths \cite{klein1993resistance,xiao2003resistance,Zhang2007,GUTIERREZ2021125751}. Although the resistance distance has been widely used for network analysis \cite{lopez2005anomalous,gallos2007scaling,leongarcia2010,bozzo2013,rubido2014resiliently,grippo2016}, its use for community detection has been limited \cite{rubido2013structure,zhang_and_bu_2019}. ZB report \cite{zhang_and_bu_2019} accurate results on small-sized networks (Zachary 's karate club \cite{Zachary1977}, dolphin social network \cite{lusseau2003emergent}, and the college football network \cite{girvan2002community}), in comparison to the methods by Kannan, Vempala, and Vetta (KVV) \cite{KVV_2001} and spectral modularity \cite{Newman8577,Newman.PRE.74.036104}. However, ZB's method lacks benchmark testing and automation.
%Ref.~\cite{zhang_and_bu_2019} shows promising results in accurate detecting the community structure of different networks. Particularly,  their results are good compared to Kannan, Vempala, and Vetta (KVV) algorithm~\cite{KVV_2001} and spectral modularity algorithm~\cite{Newman8577,Newman.PRE.74.036104}. However, these results are obtained from small real networks: Zachary 's karate club ($N=34$), dolphin social network ($N=62$) and American football club network ($N=115$); and it is necessary to define the number of communities that we wish to detect.

Here, we adapt ZB's algorithm to detect communities on networks without needing to specify the number of communities beforehand, making it an unsupervised method. We test the accuracy of our modified algorithm's detection rate in GN and LFR benchmarks, which allows us to compare it with previously reported results from other methods. Our findings show that the algorithm's accuracy is comparable -- in LFR networks -- or better -- in GN networks -- than methods such as clique percolation (also known as CPM or Cfinder) \cite{palla2005uncovering}, Markov Clustering algorithm (MCL) \cite{VanDongen}, hierarchical divisions \cite{radicchi2004defining}, and exhaustive modularity optimisation \cite{guimera2004modularity,medus2005detection}. Because the resistance distance is unrestricted to these particular networks, we expect that our algorithm can detect communities in weighted, directed, and even, evolving networks.
%Thus allowing us not to only test the method on bigger networks ($N=1000$), but also to compare our results with those already published. We found that the performance of the algorithm , i.e. the ability of the algorithm to find the community structure provided by the benchmark, is good and even better that other methods; for example, local methods of clique percolation, also known as CPM or Cfinder \cite{palla2005uncovering}, methods of random walk or difussion as the Markov Clustering algorithm (MCL) \cite{VanDongen}, some methods of hierarchy division as Radicchi's \cite{radicchi2004defining}, and algorithms of exhaustive modularity optimization \cite{guimera2004modularity,medus2005detection}.
%
%%%%%%%%%%%%%%%%%%%%%%%%%%%%%%%%%%%%%%%%%%%%%%%%%%%%%%%%%%%%%%%%%%%
%%%%%%%%%%%%%%%%%%%%%%%%%%%%%%%%%%%%%%%%%%%%%%%%%%%%%%%%%%%%%%%%%%%
\section{Methods}
%
%_________________________________________________________________%
    \subsection{Definitions and Notation}
A network is a pair of sets, $\mathcal{G} = \{\mathcal{V},\mathcal{E}\}$, where $\mathcal{V}$ is a set of $N$ nodes (vertices) and $\mathcal{E}\subset\mathcal{V}\times\mathcal{V}$ is a set of $M$ links (edges) from the unordered set of node pairs $\mathcal{V}\times\mathcal{V} = \{(i,j):\,i\in\mathcal{V},\,j\in\mathcal{V}\}$. $\mathcal{G}$ can be represented by its adjacency matrix, $\mathbf{A}$, such that $A_{ij} = 1$ if there is a link connecting node $i$ to $j$, or $A_{ij} = 0$ otherwise. When $\mathcal{G}$ is undirected and unweighted, $A_{ij} = A_{ji}$ and $N-1\leq M\leq N(N-1)/2$. Also, the node's degree is the number of neighbours of a node, such that $k_i = \sum_{i=1}^N A_{ij}$ and $\sum_{i=1}^N k_i = 2M = N\left\langle k \right\rangle$, making the network's link density be
\begin{equation}
    \rho(\mathcal{G}) = \frac{ |\mathcal{E}| }{ |\mathcal{V}\times\mathcal{V}| } = \frac{2M}{N(N-1)} = \frac{\left\langle k \right\rangle}{N-1}.
    \label{eq_link_density}
\end{equation}

A community is a subset of nodes, $\mathcal{W}\subset\mathcal{V}$, namely, a partition, that can be defined in terms of its relative link density \cite{FORTUNATO201075,reichardt2004detecting,noack2009modularity,quiles2016dynamical}, $\rho(\mathcal{W})$, such that it fulfil
\begin{equation}
    \rho(\mathcal{W}) > \rho(\mathcal{G}) > \rho(\mathcal{G}\backslash\mathcal{W}),
    \label{eq_CommunityDef}
\end{equation}
where $\rho(\mathcal{G}\backslash\mathcal{W})$ is the link density between $\mathcal{W}$ and its complement, $\bar{\mathcal{W}} = \mathcal{G}\backslash\mathcal{W}$. We note that Eq.~(\ref{eq_CommunityDef}) leads to several ways to define which nodes belong to $\mathcal{W}$, since small changes can leave it unaffected (such as, depending on the network, the effect of removing or including a node to $\mathcal{W}$). However, because of the higher density of links within a community, its nodes tend to be topologically closer.
%There is no universally accepted definition of community \cite{FORTUNATO201075,BOCCALETTI2006175}. Usually, a community is defined as a subset of nodes than is more densely connected that the average link density of the network (\ref{eq_link_density}). Moreover, it is required that the link density between this subset of nodes and the rest of the network has to be lower than the average.  Accordingly, if $W$ is a subset of nodes and $\bar{W}$ is the subset of the rest of the nodes of the network, then $W$ is a community if the following relation is verified:

A measure to quantify the topological distance between pairs of nodes is the resistance distance \cite{klein1993resistance,xiao2003resistance,Zhang2007,GUTIERREZ2021125751}. For nodes $i$ and $j$, it can be determined by
\begin{equation}
        R_{ij} = \sum_{n = 2}^N \frac{1}{\lambda_n(\mathbf{L})} \left| [\vec{\phi}_n]_i - [\vec{\phi}_n]_j \right|^2\geq0,
    \label{eq_ResistDist}
\end{equation}
where $\lambda_n(\mathbf{L})$ is the $n$-th eigenvalue of the Laplacian matrix $\mathbf{L}$ and $[\vec{\phi}_n]_i$ is the $i$-th component of the corresponding eigenvector, i.e., $\mathbf{L}\vec{\phi}_n = \lambda_n(\mathbf{L})\vec{\phi}_n$, with $\{0=\lambda_1<\lambda_2\leq\ldots\leq\lambda_N\}$. $\mathbf{L}(\mathbf{A}) = \mathbf{D} - \mathbf{A}$, where $\mathbf{D}$ is the diagonal matrix containing the node degrees, making $\mathbf{L}(\mathbf{A})$ a positive semi-defined matrix when the network is undirected. It has been shown that nodes within a community have smaller $R_{ij}$ values than nodes belonging to different communities \cite{rubido2013structure}. The reason is that $R_{ij}$ is small ($R_{ij}<1$) when there are parallel paths between nodes, but increases with serial paths ($R_{ij}>1$), as in a bridge between communities.

%The modularity $Q$ is a measure of the quality of a community structure of the network \cite{newman_comm_struct}, and it is defined as
%\begin{equation}\label{eq_modularity}
%    Q=\sum_{i=1}^{N(A)}\left\lbrace e_{ii}-\left(\sum_{j=1}^{N(A)}e_{ij}\right)^2\right\rbrace,
%\end{equation}
%where $e_{ij}$ is  $N(A)\times N(A)$ matrix which elements are the fraction of links between communities $i$ and $j$. The first term of equation $(\ref{eq_modularity})$ represents the fraction of intra-community links of the network, while the second term represents its expected -random- value. Thus, modularity has positive values, and increasingly closer to 1, if the links within communities are higher than expected \cite{Newman8577}.

%
%_________________________________________________________________%
    \subsection{Resistance-Distance-based Community Detection Method} \label{sec_algorithm}
We modify the algorithm by Zhang and Bu (ZB) \cite{zhang_and_bu_2019} to automate the partitioning process. That is, we include points 7-10 to the original algorithm.

\begin{enumerate}
    \item\textbf{Compute} $R_{ij}(p)$ from Eq.~(\ref{eq_ResistDist}) for all $N_p\times N_p$ pairs of nodes belonging to the $p$-th partition $\mathcal{V}_p\subset\mathcal{V}$.
    %The resistance distances between all $N_\ell$ nodes of the subset $V_\ell$ are calculated according to Eq.~(\ref{eq_ResistDist}). 
    
    \item\textbf{Apply} a Gaussian transformation to $R_{ij}(p)$: $S_{ij}(p) = A_{ij}\exp\left(-R_{ij}(p)^2/2\right)$, which highlights the different resistance distances between nodes.
    %This procedure transform the unweighted network $V_\ell$ into the weighted network $V_\ell'$ with adjacency matrix $\mathbf{S}$, and node's weight $\kappa_i=\sum_j^{N_\ell}s_{i,j}$. A diagonal matrix $\mathbf{D}'$ is defined, as the matrix which elements are the weight of the nodes.
    
    \item\textbf{Compute} normalised Laplacian matrix of $\mathbf{S}(p)$: $\mathcal{L}_{S(p)} = \mathbf{D}_S^{-1/2} \mathbf{L}(\mathbf{S}) \mathbf{D}_S^{-1/2}$, where $\mathbf{L}(\mathbf{S}) = \mathbf{D}_{S} - \mathbf{S}(p)$ with $[\mathbf{D}_S]_{ii} = \kappa_i = \sum_{i=1}^{N_p} S_{ij}(p)$ and $[\mathbf{D}_S]_{ij} = 0$.

    \item\textbf{Find} the eigenvector $\vec{\psi}_2$ of $\mathcal{L}_{S(p)}$ associated to the algebraic connectivity \cite{fiedler}, i.e., the smallest non-zero eigenvalue.
    
    \item\textbf{Order} $\vec{\psi}_2$ in ascending magnitude: $\vec{\psi_2} \to [\vec{\psi_2}]_{\pi_1} < [\vec{\psi_2}]_{\pi_2} < \dots < [\vec{\psi_2}]_{\pi_{N_p}}$.
    %From the indices of this ordering, that refer to nodes of the network, we obtain two vectors $W_k=\{\alpha_1,\alpha_2,\dots,\alpha_k\}$ and $\bar{W}_k=\{\alpha_{k+1},\alpha_{k+2},\dots,\alpha_{N_\ell}\}$, where $k\in \{1,2,\dots,N_\ell\}$.
    
    \item\textbf{Partition} $\mathcal{V}_p$ according to the minimisation of the Cheeger constant \cite{Chung2005LaplaciansAT}:
    %value of $k^*$ selected corresponds to the one that minimizes the Cheerger constant~\cite{Chung2005LaplaciansAT},
    \begin{equation}
        h_G(\mathcal{W}_\pi^\star) = \min_{\mathcal{W}_\pi\subset \mathcal{V}_p} \left\lbrace \frac{ \sum_{i\in \mathcal{W}_\pi} \sum_{j\in\overline{W_\pi}} A_{ij} }{ \min\left\lbrace \sum_{i\in \mathcal{W}_\pi} k_i,\sum_{i\in\overline{\mathcal{W}_k}} k_i \right\rbrace } \right\rbrace,
        \label{eq_Cheeger}
    \end{equation}
    where $\mathcal{V}_p = \mathcal{W}_\pi\cup\overline{\mathcal{W}_\pi}$ for any $\pi\in\{\pi_1,\,\pi_2,\,\ldots,\,\pi_{N_p}\}$. Equation~(\ref{eq_Cheeger}) minimises the ratio between the number of inter-partition links (numerator) and the number of intra-partition links to give the optimal $\pi^\star$, defining the subset $\mathcal{W}_\pi^\star = \{\pi_1,\,\ldots,\,\pi^\star\}$ and its complement, $\overline{\mathcal{W}_\pi^\star} = \{\pi^\star,\,\ldots,\,\pi_{N_p}\}$.
    %from which a partition of the network is obtained: $V_{\ell,1}=\{\alpha_1,\dots,\alpha_{k^*}\}$ and $V_{\ell,2}=\{\alpha_{k^*+1},\dots,\alpha_{N_\ell}\}$.
    
    \item\textbf{Update} tentative community structure: $\mathcal{B}_{new} = \{\mathcal{V}_1 \cup \cdots \cup \mathcal{V}_{p-1} \cup \mathcal{W}_\pi^\star \cup \overline{\mathcal{W}_\pi^\star} \cup \mathcal{V}_{p+1} \cup \cdots \cup \mathcal{V}_{N(B)_{new}} \} \leftarrow \mathcal{B}_{old} = \{\mathcal{V}_1 \cup \cdots \cup \mathcal{V}_{p-1} \cup \mathcal{V}_p \cup \mathcal{V}_{p+1} \cup \cdots \cup \mathcal{V}_{N(B)_{old}} \}$, where $N(B)_{new} = N(B)_{old}+1$ is the number of communities up to this point.
    %Then $B'=\{V_1\cup\dots\cup V_{\ell-1}\cup V_{\ell+1}\cup\dots\cup V_{\ell,1}\cup V_{\ell,2}\}$ is defined, with $m+1$ elements, and $I(A_{ref},B')$ is calculated according with (\ref{eq_IMN}).
    
    \item\textbf{Compute} modularity \cite{Newman8577}: $Q(\mathcal{B}_{new}) = \frac{1}{4M}\sum_{i,j}\left(A_{ij} - \frac{k_i\,k_j}{2M}\right)(s_i\,s_j)$, where $s_i\,s_j = 1$ if nodes $i$ and $j$ belong to the same community, otherwise $s_i\,s_j = 0$.
    
    \item\textbf{Accept} partition if $Q(\mathcal{B}_{new}) > Q(\mathcal{B}_{old})$. Otherwise, discard $\mathcal{B}_{new}$.
    
    \item\textbf{Repeat} steps 1 to 8 changing $p \to p+1$ while $p+1<N(B)<N$.
\end{enumerate}

We initialise the community detection by generating an initial partition, $\mathcal{B}_{init}$, by applying steps 1 to 7 on $\mathcal{G} = \mathcal{V}_0$, which divides the network into $2$ subsets based on Cheeger's constant [Eq.~(\ref{eq_Cheeger})]. This means that ZB's method can divide each subset into two new subsets, successively.
%To initialise the $B$ partition, before iterating the algorithm, steps 1-4 are run with $B=G$ and $\ell=1$ as inputs, and $B\leftarrow B'=\{V_1,V_2\}$, $I(A_{ref},B)$ and $\ell=1$ as outputs. When $\ell=n$, the proposed algorithm proceeds in the following fashion:
%
%_________________________________________________________________%
    \subsection{Method Validation: Benchmark Testing and Accuracy}
\begin{figure}[ht]
    \centering\vspace{-2pc}
    \includegraphics[width=1\columnwidth]{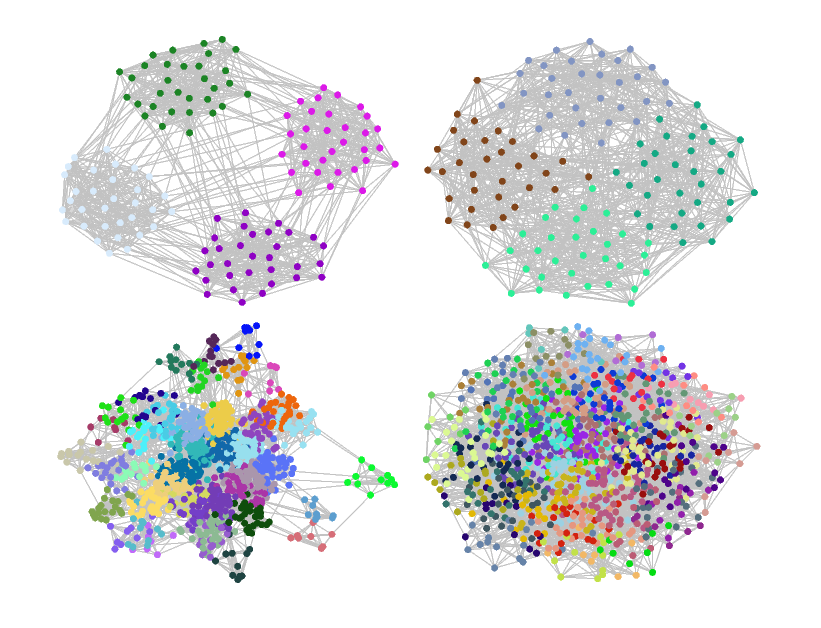}\vspace{-1pc}
    \caption{\textbf{Force-directed layouts of benchmark networks}. Top panels show Girvan-Newman \cite{girvan2002community} networks with $N=128$ nodes and average degree $\langle k \rangle = 16$. Bottom panels show Lancichinetti-Fortunato-Radicchi \cite{fortunato_benchmark} networks with $N = 1000$ nodes, average degree $\langle k \rangle = 15$, and power-law's exponent for the degree and community-size distributions being $\gamma = 2$ and $\beta = 1$, respectively. Left [Right] panels are networks with mixing parameter $\mu = 0.1$ [$\mu = 0.3$]. Colours indicate the predefined communities.}\label{fig_redes}
\end{figure}

We test our ZB algorithm's ability to correctly detect communities on $100$ network realisations of Girvan-Newman (GN) \cite{girvan2002community} and Lancichinetti-Fortunato-Radicchi (LFR) \cite{fortunato_benchmark} benchmarks. $4$ examples of these reference networks, $\{\mathcal{A}_{ref}\}$, with predefined $N(A)$ communities, are shown in Fig.~\ref{fig_redes}. We generate these benchmarks by freely available codes from Refs.~\cite{domingues_rita_code,mastersthesis_domingues} (GN) and \cite{fortunato_web} (LFR).

GN networks are generated from the Erd{\"o}s-Réyni (ER) random network model \cite{BOCCALETTI2006175}. This model sets $N(A) = 4$ communities with $n_j = 32$ nodes each ($N = \sum_{j=1}^{N(A)} n_j= 128$), assigning intra-community links with probability $p_{in}$ and inter-community links with probability $p_{out}$, where $0 < p_{out}/p_{in} < 1$. These probabilities are related to the internal and external node degrees, $k^{in}$ and $k^{out}$, respectively -- the attachment probability, $p$, in ER networks is such that $p = \rho(\mathcal{G}) = \left\langle k \right\rangle/(N-1)$ [see Eq.~(\ref{eq_link_density})]. We fix the average node degree to $\left\langle k \right\rangle = 16$ and change the mixing of communities by changing $p_{in}$ and $p_{out}$ such that
%Testing community detection algorithms involves running them on a network with a well defined community structure, and recovering the nodes' assignation to the different communities \cite{fortunato_benchmark}. Therefore, many realizations of similar networks with a know community structure are required. This is the role of the benchmark networks. Two commonly used benchmarks can be found in the literature:  the Girvan-Newman (GN) benchmark \cite{girvan2002community} and the Lancichinetti-Fortunato-Radicchi (LFR) benchmark \cite{fortunato_benchmark}. 
%GN networks follow the Erd{\"o}s-Réyni model: a link between two nodes exists with probability $p_{in}$, if both links belong to the same community, or with probability $p_{out}$ if they belong to different communities. It must be verified that $p_{out}<p_{in}$ in order to obtain a community structure \cite{girvan2002community}. These probabilities are related to the internal and external degree ($k^{in}$ and $k^{out}$) of the nodes, i.e. the number of links connecting the node with a different community, and the number of links connecting the node with other nodes of the same community, respectively. These benchmark networks are usually set with 128 nodes split in 4 different communities of 32 nodes each. Although the average degree of the network can be changed, in this work we only considered $\langle k\rangle=16$. Hence, the only control parameter that we varied in this class of networks was the mixing parameter, $\mu_i$, defined as
\begin{equation}\label{eq_mixing}
    \mu_i = \frac{ k_i^{out} }{ k_i^{out}+k_i^{in} } < 1.
\end{equation}
The mixing control parameter is then the network average $\sum_i \mu_i/N = \mu \in[0,1]$.

LFR networks are generated by assigning each node degree, $k_i$, from a power-law distribution: $P(k)\sim k^{-\gamma}$. The distribution range is set to satisfy the prefixed network-average, $\langle k \rangle$. The size of each community, $n_j$, is assigned from another power-law distribution: $P(n)\sim n^{-\beta}$, such that $\sum_{j=1}^{N(A)} n_j = N$. Then, nodes are assigned randomly to these communities, as long as the community size is bigger than the internal degree of the node. Finally, links are assigned by various rewiring steps that modify the internal and external degrees of each node, without modifying the node's degree, $k_i$, until $k^{in}\approx\mu k_i$ and $k^{out}\approx(1-\mu)k_i$, where the mixing parameter is defined by Eq.~(\ref{eq_mixing}).

%%%_________________________________________________________________
%%% ESTO VA EN LOS RESULTADOS O EN LA DISCUSION DE NUESTROS RESULTADOS, PARA PONERLOS EN CONTEXTO CON LO QUE SE HA ENCONTRADO CON OTROS ALGORITMOS
%Also on GN and LFR networks, Fortunato et al. test different community detection algorithms that can be found in the literature \cite{fortunato2009community}. From their results, they classify the algorithms into three categories: bad algorithms, which performance rapidly decline with the increase of the mixing parameter; fair algorithms, which performance also decline with the increase of the mixing parameter, but not as fast; and good algorithms, which can faithfully discover the community structure of the network for $\mu\lesssim 0.5$.

We measure the community detection accuracy by the Normalised Mutual Information, $I$ \cite{danon2005comparing,gouvea2021force}, which quantifies the similarity between a reference structure, $\mathcal{A}_{ref} = \{\mathcal{V}_1 \cup \mathcal{V}_2 \cup \ldots \cup \mathcal{V}_{N(A)}\}$ with $N(A)$ communities, and a detected structure, $\mathcal{B} = \{\mathcal{W}_1 \cup \mathcal{W}_2 \cup \ldots \cup \mathcal{W}_{N(B)}\}$ with $N(B)$ communities. Specifically,
%Given a network $G$, a reference partition $A_{ref}$ of $G$ (the one given by the network's generation code), $B=\{V_1\cup V_2\cup\dots\}$ another partition of $G$ with $m$ elements, $\ell$ an index that runs through the elements of $B$, and $I(A_{ref},B)$ the normalized mutual information defined as \cite{danon2005comparing}
\begin{equation}
    I(\mathcal{A}_{ref},\mathcal{B}) = \frac{ -2 \sum_{i=1}^{N(A)} \sum_{j=1}^{N(B)} C_{ij}\log\left(\frac{C_{ij}\,N}{C_i\,C_j}\right) }{ \sum_{i=1}^{N(A)} C_i\log\left( \frac{C_i}{N} \right) + \sum_{j=1}^{N(B)} C_j\log\left( \frac{C_j}{N} \right) },
    \label{eq_IMN}
\end{equation}
where $C_{ij}$ is the number of nodes in community $i$ of $\mathcal{A}_{ref}$ that are also in community $j$ of $\mathcal{B}$, $C_i = \sum_j C_{ij}$, and $C_j = \sum_i C_{ij}$.

%
%%%%%%%%%%%%%%%%%%%%%%%%%%%%%%%%%%%%%%%%%%%%%%%%%%%%%%%%%%%%%%%%%%%
%%%%%%%%%%%%%%%%%%%%%%%%%%%%%%%%%%%%%%%%%%%%%%%%%%%%%%%%%%%%%%%%%%%
\section{Results}

\begin{figure}[ht]
    \centering
    \includegraphics[width=1.0\columnwidth]{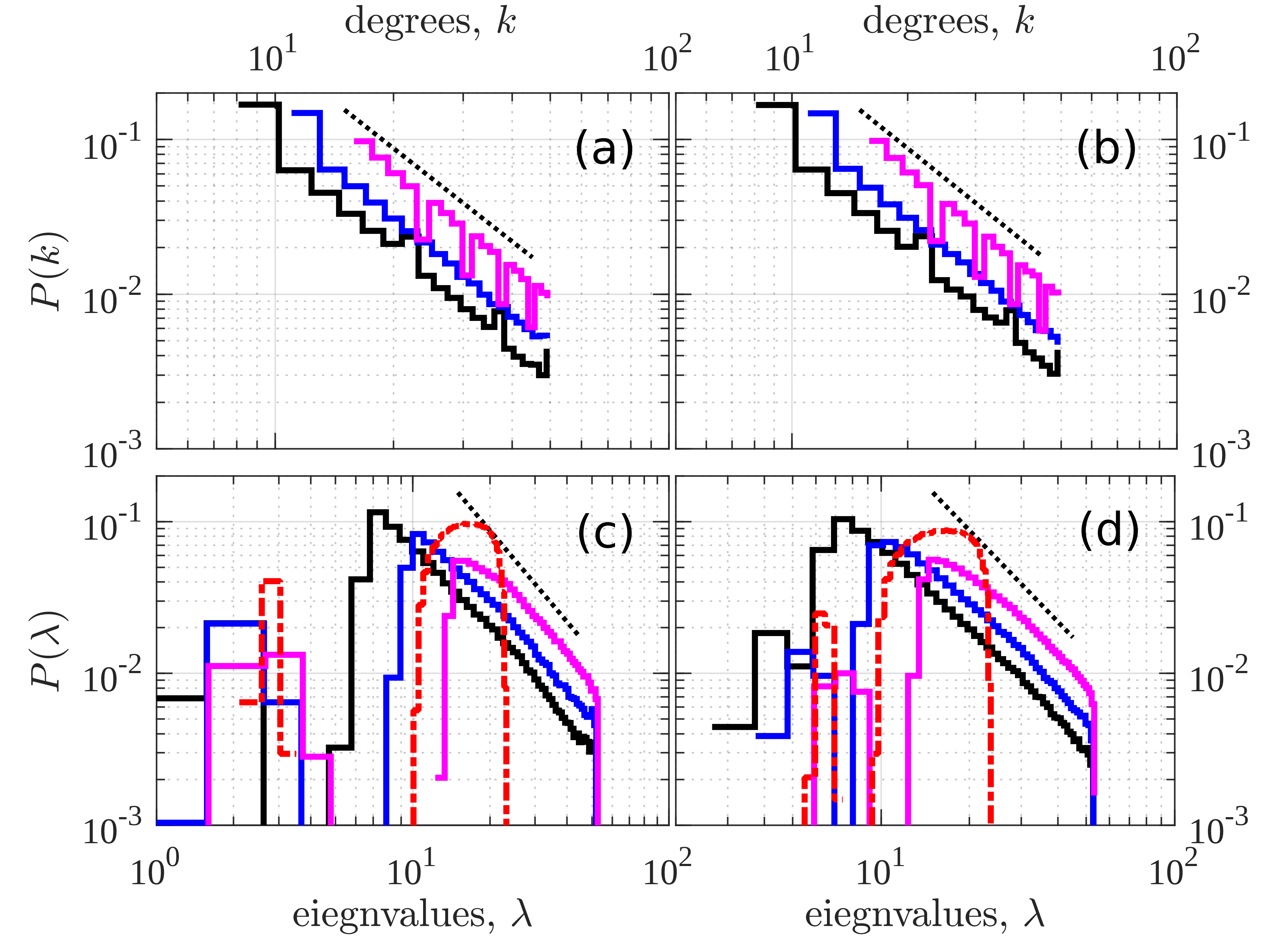}
    \caption{\textbf{Degree and spectral probability distributions, $P(k)$ and $P(\lambda)$, of networks with communities}. All panels have Lancichinetti-Fortunato-Radicchi networks \cite{fortunato_benchmark} with $N = 1000$ nodes, generated from $P(k) \sim k^{-2}$ (dotted line) and communities with sizes following $P(n) \sim n^{-1}$, where $\langle k \rangle = 15$ (black), $\langle k\rangle = 20$ (blue), or $\langle k \rangle = 25$ (magenta), changing the distribution ranges. Bottom panels have Girvan-Newman networks \cite{girvan2002community} with $N =128$ and $\langle k \rangle = 16$ (red dashed). Panels {\bf (a)} and {\bf(c)} [{\bf (b)} and {\bf(d)}] show the distributions when the mixing parameter $\mu = 0.1$ [$\mu = 0.3$].}% Top panels show the degree distributions (not shown for GN networks since all nodes have almost the same degree for this class of networks) and bottom panels show the spectral distribution of the Laplacian matrices, $\mathbf{L}$. Dashed black line represents a power law with exponent 2.}
    \label{fig_dist_unweight}
\end{figure}

Figure~\ref{fig_dist_unweight} shows degree (top panels) and spectral (bottom panels) distributions, $P(k)$ and $P(\lambda)$, respectively, for $8$ benchmark networks. The $6 $LFR networks show power-law degree-distributions, which fit the expected exponent $\gamma = 2$, as can be seen by the dotted lines in Figs.~\ref{fig_dist_unweight}{\bf (a)} and {\bf (b)}. Because the Laplacian matrix is closely related to the node degrees, the spectral distributions, $P(\lambda)$, also tend to a power-law with similar exponent, as can be seen from their tails in Figs.~\ref{fig_dist_unweight}{\bf (c)} and {\bf (d)}. On the other hand, GN networks have a narrow degree distribution (not shown), since we fix $\langle k \rangle = 16$, which implies that $P(k) \sim \delta(k-16)$. Because of the random features, GN spectral distribution tends to Wigner's semi-circle distribution, as can be seen from the dashed curves in Fig.~\ref{fig_dist_unweight}{\bf (c)} and {\bf (d)}. These characteristics are translated to the resistance distance by means of Eq.~(\ref{eq_ResistDist}), where the smaller eigenvalues have the larger influence, particularly, the algebraic connectivity (also known as Fiedler eigenvalue \cite{fiedler}).
%are shown for some of the networks analysed in this work. We can see the power laws that both follow on LFR networks, both with the same exponent. Degree distributions corresponding to GN networks are not shown, since these are very narrow, i.e. a Dirac delta function.

\begin{figure}[ht]
    \centering
    \includegraphics[width=1.0\columnwidth]{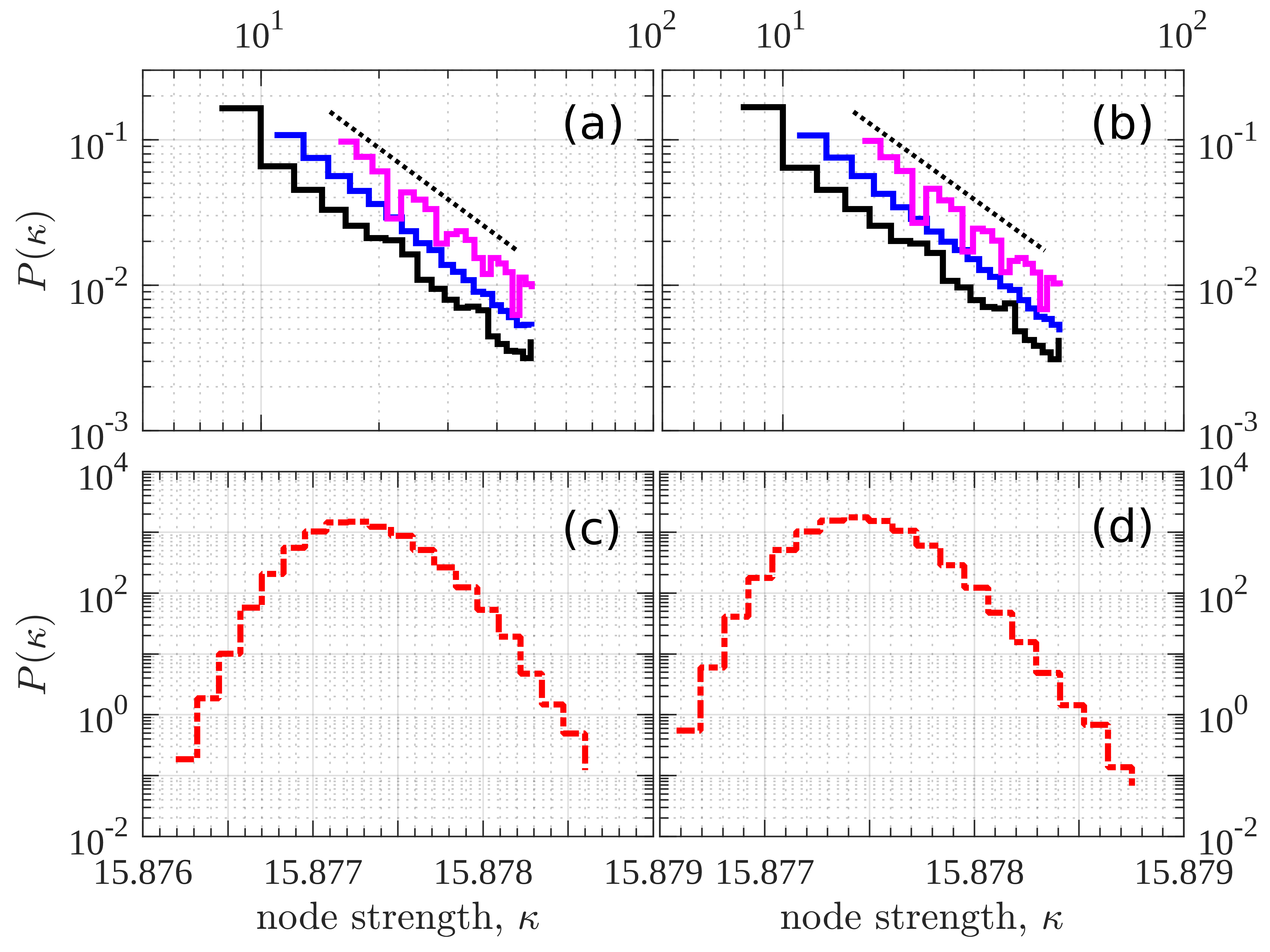}
    \caption{\textbf{Node strength probability distributions, $P(\kappa)$, of benchmark networks}. The $i$-th node strength, $\kappa_i = \sum_j S_{ij}(R_{ij})$, is the weighted-degree of the Gaussian-transformed resistance distance, $R_{ij}$, of the network [Eq.~(\ref{eq_ResistDist})]. Top panels show $P(\kappa)$ for the Lancichinetti-Fortunato-Radicchi networks of Figs.~\ref{fig_dist_unweight}{\bf (a)} and {\bf (b)}. Bottom panels show $P(\kappa)$ for the Girvan-Newman networks in Fig.~\ref{fig_redes}. Mixing parameters, colours, and symbols follow those of Fig.~\ref{fig_dist_unweight}.}
    %In all panels green refers to $\langle k\rangle=15$, magenta to $\langle k\rangle=20$ and blue to $\langle k\rangle=25$, corresponding to LFR networks of $1000$ nodes, $\gamma=2$ and $\beta=1$, and red refers to GN networks with $\langle k\rangle=16$. All panels show the nodes' weight distribution, $\kappa_i=\sum_j^Ns_{i,j}$, top panes in logarithmic scale and in bottom panels in horizontal lineal scale. Dashed black line represents a power law with exponent 2.}
    \label{fig_dist_weight}
\end{figure}

As can be seen from Fig.~\ref{fig_dist_weight}, the shape of the degree distributions from Fig.~\ref{fig_dist_unweight} remain nearly unaltered when analysing node strengths, $\kappa$, resultant from the weighted degrees of the Gaussian transformation of the resistance distance (steps 2 ad 3 in Sect.~\ref{sec_algorithm}). Specifically, we find power-law distributions, $P(\kappa)\sim\kappa^{-2}$, for the LFR networks -- Figs.~\ref{fig_dist_weight}{\bf (a)} and {\bf (b)} -- and a narrowly Gaussian-like distribution for the GN networks -- Figs.~\ref{fig_dist_weight}{\bf (c)} and {\bf (d)}. These distributions hold for both mixing parameters, $\mu = 0.1$ and $\mu = 0.3$, which also correspond to the networks shown in Fig.~\ref{fig_redes}. Because the node strength distribution share similarities with the node degree distribution, we also find that the spectral distribution of the Gaussian-transformed resistance distance is similar to that of the initially unweighted network (not shown). This means that ZB's community detection method (Sect.~\ref{sec_algorithm}) keeps the main topological properties of the original network and that its partitioning process can be narrowed to the network's Fiedler eigenvector, which fits into the category of spectral methods.
%we show how the degree distribution transforms into the nodes' weight distribution when we go from the original unweighted network, $G$, to the weighed network, $G'$. We observe that for LFR networks the distribution of the nodes' weight continue to follow a power law with the same exponent, while for GN networks the Dirac delta distribution has transformed in a very sharp Gaussian distribution. That the degree distributions of the networks $G$ and those of the nodes' weight are related is due that matrix $\mathbf{S}$, that represents network $G'$, is calculated from $\mathbf{L}$.

\begin{figure}[ht]
    \centering
    \includegraphics[width=1\columnwidth]{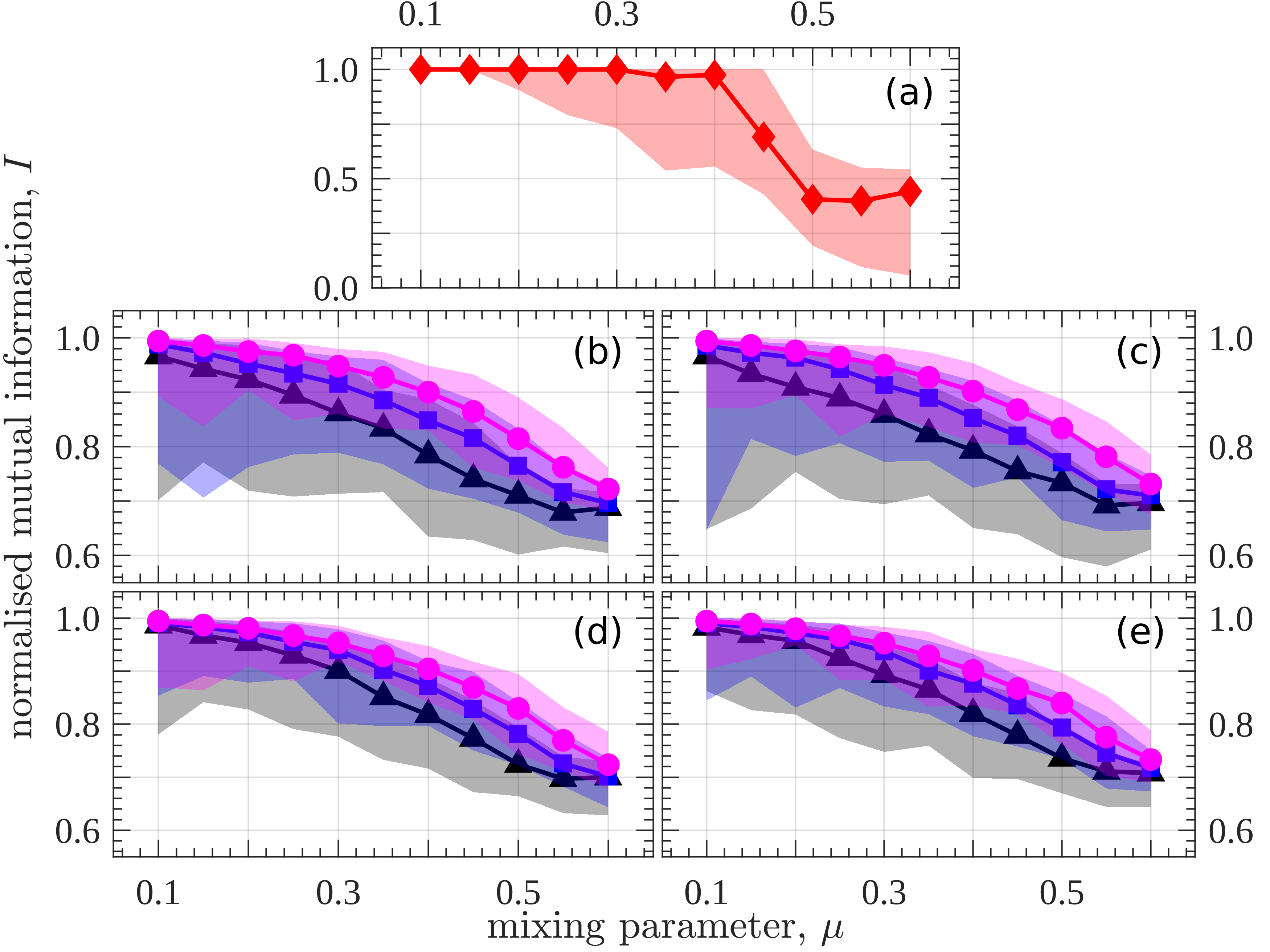}
    \caption{\textbf{Resistance-distance based community-detection accuracy as the communities becomes mixed}. Accuracy is measured by the normalised mutual information, $I$ [Eq.~(\ref{eq_IMN})], and community mixing is controlled by $\mu$ [Eq.~(\ref{eq_mixing})]. Panel {\bf (a)} show resultant detection accuracy from $100$ GN network realisations, where $N = 128$ nodes and $\langle k \rangle = 16$ average degree. Remaining panels show resultant $I$ from $100$ LFR network realisations, where $N = 10^3$ and $\langle k \rangle = 15$ (black), $\langle k\rangle = 20$ (blue), or $\langle k\rangle=25$ (magenta), as in Figs.~\ref{fig_dist_unweight} and \ref{fig_dist_weight}. The power-law exponents, $\gamma$ and $\beta$, for the LFR networks' node-degree and community-size are $(\gamma,\,\beta) = (2,\,1)$ in panel {\bf (b)}, $(\gamma,\,\beta) = (2,\,2)$ in panel {\bf (c)}, $(\gamma,\,\beta) = (3,\,1)$ in panel {\bf (d)}, and $(\gamma,\,\beta) = (3,\,2)$ in panel {\bf (e)}. Every symbol represents the median $I$ of $100$ realisations and shaded areas the $98\%$ central $I$ values.}
    %for different configurations of LFR ($N=1000$) and GN ($N=128$ and $k=16$) networks. In panel {\bf (a)} the results obtained for the LFR benchmark set with $(\gamma,\beta)=(2,1)$, in {\bf (b)} with $(\gamma,\beta)=(2,2)$, in {\bf (c)} with $(\gamma,\beta)=(3,1)$ and in panel {\bf (d)} with $(\gamma,\beta)=(3,2)$. In panel {\bf (e)} the results obtained for the GN benchmark are showed. Colors represent the different average degree of the network: in green for $\langle k\rangle=15$, in magenta for $\langle k\rangle=20$, in black for $\langle k\rangle=25$, and in red for $\langle k\rangle=16$. Every data point represents the statistical distribution of 100 realisations, with the diamonds representing the median with the $98\%$  central quantile.}
    \label{fig_N1000}
\end{figure}

In spite of ZB's method being (in its core) a spectral partitioning method, the inclusion of the resistance distance [Eq.~(\ref{eq_ResistDist})] as the matrix to use for the partitioning process results in more accurate community detection -- as we show in Fig.~\ref{fig_N1000}. In particular, the normalised mutual information, $I$, values we obtain for $100$ GN network realisations show excellent results ($I\simeq1$) up to $\mu = 0.4$, when $I$ starts to decline -- this can be seen from Fig.~\ref{fig_N1000}{\bf (a)}. On the other hand, results for $100$ LFR network realisations show a monotonous decline in accuracy, from $I \simeq 1$ when $\mu = 0.1$ to $I \simeq 0.7$ when $\mu = 0.6$, as can be seen from the remaining panels in Fig.~\ref{fig_N1000}. These values are nearly unchanged when considering different network-average node degrees, $\langle k \rangle$, which are shown by differently coloured curves and symbols (following the same pattern as in Figs.~\ref{fig_dist_unweight} and \ref{fig_dist_weight}). However, in all LFR tests, $I$ slightly increases for all $\mu$ when $\langle k \rangle$ is increased. Overall, we note that the algorithm perform sufficiently accurate, particularly, performing better than previously reported results \cite{quiles2016dynamical,fortunato_benchmark,fortunato2009community} that use Fast Greedy Optimisation \cite{clauset2004finding} or Label Propagation \cite{raghavan2007near}, to name a few.

%Different configurations of these networks were considered. For LFR networks, every realisation was set with same number of nodes, $N=1000$, varying the exponents of the degree distribution, $\gamma$, and of the communities size distribution, $\beta$, in the following values: $(\gamma,\beta)=\{(2,1),(2,2),(3,1),(3,2)\}$. In each case, two other parameters were varied: the average degree of the network, $\langle k\rangle$, in the values $\langle k\rangle=15,20,25$; and the mixing parameter, $\mu$. For GN networks, every realisations was set with $N=128$, $\langle k\rangle=16$ and $4$ communities, varying only the mixing parameter $\mu$. For every case, $100$ realisations were performed, showing the statistical distribution of the results: the diamonds representing the median with the $98\%$ central quantile.

In this accuracy analysis we have also used different LFR parameters to explore the effect of changing the degree distribution exponent, $\gamma$, and community size distribution exponent, $\beta$. For example, when $\gamma = 2$ degree distributions are as in Fig.~\ref{fig_dist_unweight}, where there are some hub nodes (given by the distribution tail). The resultant accuracy for these networks can be seen in Figs.~\ref{fig_N1000}{\bf (b)} and {\bf (c)}, whose difference comes from having more ($\beta = 1$) or fewer ($\beta = 2$) communities with varying size, respectively. Similarly, we analyse the resultant detection accuracy when $\gamma = 3$, where the number of hubs decreases but they increase their degrees. Resultant $I$ values are shown in Figs.~\ref{fig_N1000}{\bf (d)} and {\bf (e)}, where differences emerge from changing $\beta$ as in panels {\bf (b)} and {\bf (c)}. Overall, Fig.~\ref{fig_N1000} shows that ZB's method for community detection works fairly on GN and LFR networks, complementing their previous results on small-sized real-world networks \cite{zhang_and_bu_2019}.
%
%%%%%%%%%%%%%%%%%%%%%%%%%%%%%%%%%%%%%%%%%%%%%%%%%%%%%%%%%%%%%%%%%%%
%%%%%%%%%%%%%%%%%%%%%%%%%%%%%%%%%%%%%%%%%%%%%%%%%%%%%%%%%%%%%%%%%%%
\section{Discussion and Conclusions}
%%%__________________________________________________________________
%%% HASTA ACA LLEGUE HOY
Our works is based on extending Zhang and Bu (ZB) \cite{zhang_and_bu_2019} method for community detection to automate its operation and quantify its accuracy to correctly detect communities in benchmark networks. In order to successively partition the network into smaller modules, ZB method follows Kannan-Vempala-Vetta (KVV) bi-sectioning algorithm \cite{KVV_2001} (steps 2-6 in Sect.~\ref{sec_algorithm}), but it uses the resistance distance of the network (step 2 in Sect.~\ref{sec_algorithm}) instead of its adjacency matrix. We add modularity optimisation to the process (steps 7-10), which makes the resultant algorithm a hybrid method involving resistance distance, spectral partitioning, and modularity optimisation. Consequently, our adaptation allows to iterate the algorithm without needing to specify the number of communities in the network or control its outcomes, making it an unsupervised algorithm.
%Steps (ii)-(iv) of the algorithm here presented correspond to the KVV bisection algorithm \cite{KVV_2001}. Zhang-Bu implementation of this algorithm for community detection correspond to step (i), where the resistance distance is used to construct a similarity measure. 
%Steps (v)-(vii), constitute our adaptation of the method to be able to iterate the algorithm without the need of previously stating the number of communities we want to detect, finishing when the discovery of new communities does not increase the normalised mutual information. In this work, we used the normalised mutual information as a measure to decide if a new partition of the network represents a possible community structure of the network.

In order to quantify the method's accuracy, we use Girvan-Newman (GN) and Lancichinetti-Fortunato-Radicchi (LFR) benchmark networks, where $N = 128$ and $1000$, respectively. These classes of networks are different in size and overall topology. For example, the mixing parameter, $\mu$, in GN networks relates to the probabilities of intra- and inter-community links. On the other hand, $\mu$ in LFR networks relates to the rewiring process by which scale-free degrees and communities are inter-connected. However, according to our results from Fig.~\ref{fig_N1000}, the modified ZB algorithm can detect communities in networks with significant mixing, such as $\mu = 0.4$ or $0.5$, which as can be seen from the force-directed layout in Fig.~\ref{fig_redes}, the community structure presents serious challenges.

These benchmark tests extend ZB results (where $N\lesssim115$) and complements their work. In particular, Fortunato et al. \cite{fortunato2009community} classifies community detection algorithms as a function of $\mu$ and their performance on benchmark networks according to $3$ categories: a) bad (those that $I\to0$ rapidly with increasing $\mu$), b) fair (those that $I$ declines with increasing $\mu$ but remains finite), and c) good (those where $I\sim1$ for $\mu\lesssim0.5$). Therefore, ZB's algorithm -- with our modifications -- falls into the fair category on LFR networks and good category on the GN networks, as it can be corroborated from Fig.~\ref{fig_N1000}. Although, we note that there is room for improvement in terms of its the computational complexity, which is the main drawback in most spectral methods.
%We based this performance testing in the comparative study presented in Ref.~\cite{fortunato2009community}, which allowed us to compare the obtained results with the performance of other algorithms commonly used in the literature. In this line, we observe that the ability of the proposed algorithm to detect the community structure is good, comparable to the best results reported for Girvan-Newman, and to the fair performance algorithms for LFR benchmark. We base this affirmation in the following observations: $I>0.8$ for $\mu\lesssim0.4$ for all cases considered, and there is no region where $I\rightarrow0$, particularly, it is never observed that $I\rightarrow0$ when $\mu\rightarrow0.6$ in any case. 
%
%%%%%%%%%%%%%%%%%%%%%%%%%%%%%%%%%%%%%%%%%%%%%%%%%%%%%%%%%%%%%%%%%%%
%%%%%%%%%%%%%%%%%%%%%%%%%%%%%%%%%%%%%%%%%%%%%%%%%%%%%%%%%%%%%%%%%%%
%\section*{References}
\bibliographystyle{unsrt}
\bibliography{bibliografia}

\end{document}